# Equations of Motion of Systems with Internal Angular Momentum - II

M. Dorado
CiRTA, SL. Pol. Ind. Tres Cantos Oeste, s/n

**Abstract**
Using the Euler's equations and the Hamiltonian formulation, an attempt has been made to obtain the equations of motion of systems with internal angular momentum that are moving with respect to a reference frame when subjected to an interaction. This interaction involves the application of a torque that is permanently perpendicular to the internal angular momentum vector.

## INTRODUCTION

The angular momentum of a many-particle system with respect to its centre of mass is known as the "internal angular momentum" and is a property of the system that is independent of the observer. Internal angular momentum is therefore and attribute that characterizes a system in the same way as its mass or charge. In the case of a rigid body and particularly in the case of an elementary particle, the internal angular momentum is also referred to as "spin".

To the author's knowledge, the dynamical behaviour of systems with internal angular momentum has not been systematically studied within the framework of classical mechanics (See References [1]). This paper approaches this type of problem through a model comprising a rotating cylinder, with constant velocity of rotation, around its longitudinal axis. The equations of motion are obtained using the Euler's equations and by the Hamiltonian procedure.

Results coincide when the problem is solved using vectorial algebra or Lagrangian formalism (unpublished). However, as a result of changes in "perspective", each method uncovers new peculiarities regarding the intimate nature of the system's behaviour.

## FORMULATION OF THE PROBLEM

Let us consider a rigid cylindrical solid with internal angular momentum (with respect to its centre of mass) $\vec{L}$, whose centre of mass moves at a constant velocity $\vec{\nu}$ with respect to a reference frame which can be defined as soon as the interaction in the cylinder starts. It is aimed to obtain the equations of motion to describe the dynamical behaviour of the system from the instant that it undergoes an interaction, by applying a torque $\vec{M}$ that is permanently perpendicular to the internal angular momentum (See Fig.1).

When solving the problem, the following points are taken into account:

(a) The cylinder will continue spinning about its longitudinal axis at a constant angular velocity $\vec{\omega}$ throughout all the movement. In other words, its internal angular momentum module is constant. The energy that the cylinder possesses as a result of its rotation about its longitudinal axis is consequently considered to be internal and does not interfere with its dynamical behaviour.

(b) The derivative of the internal angular momentum, $\vec{L}$, with respect to a frame of axes of inertial reference ($X$, $Y$, $Z$) satisfies the equation

$$\left(\frac{d\vec{L}}{dt}\right)_{XYZ} = \left(\frac{d\vec{L}}{dt}\right)_{X'Y'Z'} + \vec{\Omega} \times \vec{L} \qquad (1)$$

in which $\vec{\Omega}$ is the rotation velocity of the frame linked to the solid ($X'$, $Y'$, $Z'$) about the frame of inertial reference axes ($X$, $Y$, $Z$).

(c) Only infinitesimal motions are considered that are compatible with the system's configurational variations brought about by the applied torque.

Email: mdorado@cirta.es

(d) The total energy of the system is not explicitly dependent on time.

(e) No term is included that refers to the potential energy. According to hypothesis, the applied torque is a null force acting on the system and the possibility of including a potential from which the applied torque derives is unknown.

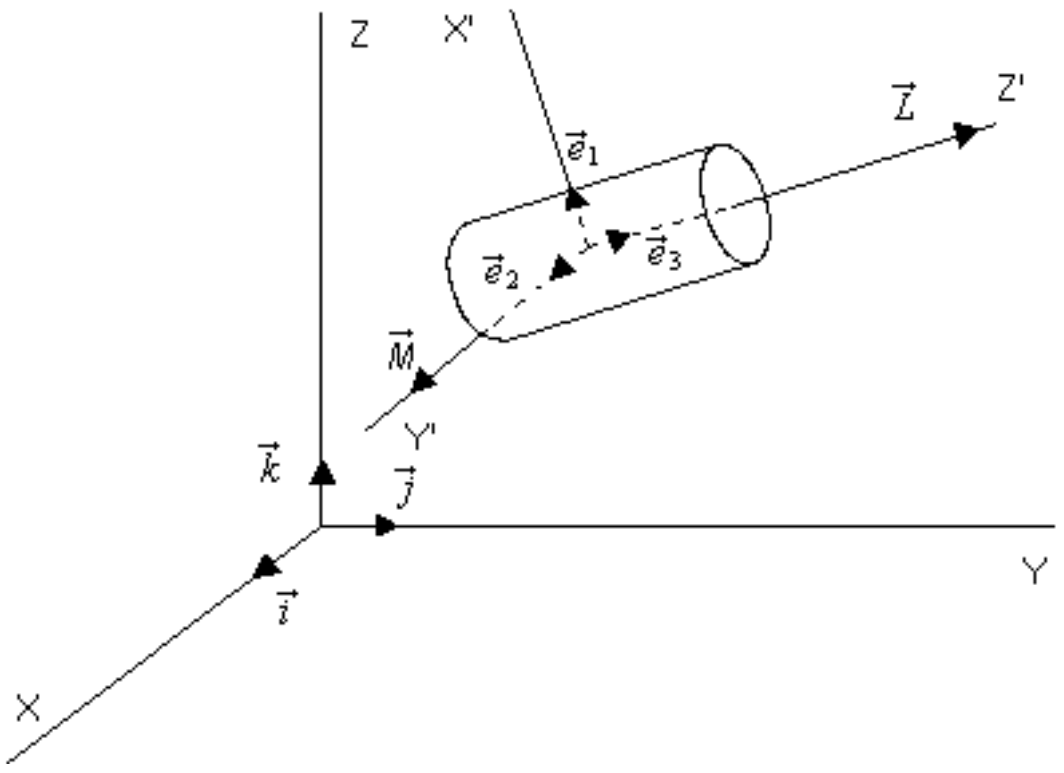

Figure 1. Formulation of the problem

## HOW TO APPROACH THESE PROBLEMS THROUGH THE EULER'S EQUATIONS?

The figure 1 shows a cylindrical rigid body with angular momentum, $\vec{L}$, about its longest axis. We should consider two referential frames: one associated with the solid ($X'$, $Y'$, $Z'$), having the $Z'$ axis on the direction of the body's longest axis. We will refer this system of coordinates to a inertial frame ($X$, $Y$, $Z$).

In a given instant, a torque $\vec{M}$ is applied on the rigid body.

We call $\vec{\Omega}$ to the rotational velocity of the system of coordinates associated with the solid ($X'$, $Y'$, $Z'$), viewed from the inertial frame ($X$, $Y$, $Z$). From here, we will consider the system of coordinates ($X'$, $Y'$, $Z'$) all along the work.

Assuming this situation, it is known that

$$\left(\frac{d\vec{L}}{dt}\right)_{XYZ} = \left(\frac{d\vec{L}}{dt}\right)_{X'Y'Z'} + \vec{\Omega}\times\vec{L} \qquad (2)$$

It is clear that we have two different velocities acting:

(a) $\vec{\omega}$: the rotational velocity of the body around its longest axis (let us call it the intrinsic rotational velocity of the body).

(b) $\vec{\Omega}$: the rotational velocity of the system of coordinates associated with the body ($X'$, $Y'$, $Z'$) viewed from the frame of inertia ($X$, $Y$, $Z$).

We can write $\vec{L}$ and $\vec{M}$, refered to ($X'$, $Y'$, $Z'$), as follows:

$$\vec{L} = I_1\omega_1\vec{e}_1 + I_2\omega_2\vec{e}_2 + I_3\omega_3\vec{e}_3 \qquad (3)$$
$$\vec{\Omega} = \Omega_1\vec{e}_1 + \Omega_2\vec{e}_2 + \Omega_3\vec{e}_3 \qquad (4)$$
$$\vec{M} = M_1\vec{e}_1 + M_2\vec{e}_2 + M_3\vec{e}_3 \qquad (5)$$

So, the equation (2) takes the expression

$$\begin{aligned} I_1\dot{\omega}_1 + I_3\omega_3\Omega_2 - I_2\omega_2\Omega_3 &= M_1 \\ I_2\dot{\omega}_2 + I_1\omega_1\Omega_3 - I_3\omega_3\Omega_1 &= M_2 \\ I_3\dot{\omega}_3 + I_2\omega_2\Omega_1 - I_1\omega_1\Omega_2 &= M_3 \end{aligned} \qquad (6)$$

These equations are known as *modified Euler's equations* and can also be found in (1.e).

In the case treated above, we must substitute the angular momentum

$$\vec{L} = I_3\omega_3\vec{e}_3 \qquad (7)$$

in (6):

$$\begin{aligned} I_3\omega_3\Omega_2 &= M_1 \\ I_3\omega_3\Omega_1 &= M_2 \\ I_3\dot{\omega}_3 &= M_3 \end{aligned} \qquad (8)$$

And then we get

$$\begin{aligned} \Omega_2 &= \frac{M_1}{I_3\omega_3} \\ \Omega_1 &= \frac{M_2}{I_3\omega_3} \\ \dot{\omega}_3 &= \frac{M_3}{I_3} \end{aligned} \qquad (9)$$

If the torque is on the axis $Y'$, $M_1 = 0$, $M_3 = 0$ and:

$$\begin{aligned} \Omega_2 &= 0 \\ \Omega_1 &= \frac{M_2}{I_3\omega_3} \\ \dot{\omega}_3 &= 0 \end{aligned} \qquad (10)$$

Where $\Omega_1$ coincides with the known velocity of precession.

We have obtained that the body moves along a circular trajectory with a rotational velocity $\Omega_1$, but this is not enough to determinate the radius of the orbit.

The principle of conservation for the energy should lead us to the expression for this radius.

Assuming our initial hypothesis, the forces are applied perpendicular to the plane of movement of the solid and so, these forces do not produce any work while the cylinder is moving.

We have two conditions that have to be satisfied by the body while rotating:

(a) $\Omega = \frac{M}{L}$

(b) The total energy should remain constant.

Before the torque is applied the energy of the cylinder takes the following expression:

$$E = \frac{1}{2}m\nu_o^2 + \frac{1}{2}I\omega_0^2 \tag{11}$$

When the torque is acting on the cylinder about the line defined by the unit vector $\vec{e}_2$, it rotates along a trajectory, generally defined by $r(t)$ and $\dot{\theta}(t)$ that, in this case, coincides with $\Omega(t)$ and its energy takes the following expression:

$$E = \frac{1}{2}m\left(\dot{r}^2 + r^2\dot{\theta}^2\right) + \frac{1}{2}I\omega_0^2 \tag{12}$$

In this particular case our axes are principal axes of inertia, therefore we have $\frac{\delta L_{X'}}{\delta t} = M_1$, and then, $\frac{\delta}{\delta t}(mr^2\dot{\theta}) = 0$.

And therefore $mr^2\dot{\theta} = const.$

In the particular case in which $\dot{\theta} = \Omega$ is a constant, $r$ has to be a constant (that is, $\dot{r}$ has to be zero) in order to satisfy this equation. We can conclude that the body moves in a circular orbit. The energy takes the following expression:

$$E = \frac{1}{2}mr^2\Omega^2 + \frac{1}{2}I\omega_0^2 \tag{13}$$

Both expressions (11) and (13) have to represent the same energy. Therefore, comparing the two expressions we can obtain the radius of the orbit:

$$r = \frac{\nu_0}{\Omega} \tag{14}$$

Looking at the results arising from the previous study, we can say that the body rotates following a circular orbit, with a radius given by the last expression. This circular trajectory implies that the velocity vector must precess jointly with the intrinsic angular momentum vector. From this last conclusion it can be proved that the force needed to cause the particle to draw a circular trajectory is expressed as:

$$\vec{F} = m\vec{\nu} \times \vec{\Omega} \tag{15}$$

When the internal angular moment of the system does not coincide with one of the principal axes of inertia, the treatment of the problem is much more complex, but as it will be seen in the next section, the results given above area completely general.

## HAMILTONIAN FORMULATION. EQUATIONS OF MOTION AND THEIR SOLUTION

The equations of motion are obtained by the Hamiltonian formulation, where the independent variables are the generalized coordinates and moments. To do this, a basis change of the frame $(q,\, \dot{q},\, t)$ to $(p,\, q,\, t)$ is made using the Legendre transformation.

From the function

$$H(p,q,t) = \sum_i \dot{q}_i p_i - L(q,\dot{q},t) \tag{16}$$

a system of $2n + 1$ equations is obtained.

$$\dot{q}_i = \frac{\partial H}{\partial p_i}; \qquad -\dot{p}_i = \frac{\partial H}{\partial q_i}; \qquad -\frac{\partial L}{\partial t} = \frac{\partial H}{\partial t} \tag{17}$$

If the last equation in (17) is excluded, a system of $2n$ first order equations is obtained, known as canonical Hamilton equations.

By substituting $H$, the first order equations of motion, are obtained.

It should be noted that the Hamiltonian formulation is developed for holonomous systems and forces derived from a potential that depends on the position or from generalized potentials. A torque is applied to the present system. The result of the forces is zero on the centre of mass and, therefore, it is meaningless to refer to potential energy.

On the other hand, the Hamiltonian function concept does remain meaningful.

By using polar coordinate in the movement plane,

$$\nu_r = \dot{r}; \qquad v_\theta = r\dot{\theta} \tag{18}$$

The Lagrangian is

$$L = \frac{1}{2}m(\dot{r}^2 + r^2\dot{\theta}^2) \tag{19}$$

The generalized moments are

$$P_r = m\dot{r}; \qquad P_\theta = mr^2\dot{\theta} \tag{20}$$

so that

$$\dot{r} = \frac{P_r}{m}; \qquad \dot{\theta} = \frac{P_\theta}{mr^2} \tag{21}$$

The Hamiltonian $H$ is introduced using the equation

$$H(p,q,t) = \sum_i \dot{q}_i p_i - L(q,\dot{q},t) \tag{22}$$

resulting in

$$H = P_r\dot{r} + P_\theta\dot{\theta} - \left(\frac{1}{2}m\dot{r}^2 + \frac{1}{2}mr^2\dot{\theta}^2\right)$$

If the generalized velocities are substituted by the generalized moments, the following is obtained

$$H = \frac{P_r^2}{2m} + \frac{P_\theta^2}{2mr^2} \tag{23}$$

The first pair of Hamilton equations is

$$\dot{r} = \frac{\partial H}{\partial P_r} = \frac{P_r}{m}; \qquad \dot{\theta} = \frac{\partial H}{\partial P_\theta} = \frac{P_\theta}{mr^2}$$

The second pair of equations is

$$-\dot{P}_r = \frac{\partial H}{\partial r} = -\frac{P_\theta^2}{mr^3}; \qquad -\dot{P}_\theta = \frac{\partial H}{\partial \theta} = 0$$

The second of these equations shows that the angular momentum $J$ is conserved.

$$P_\theta = J = const. \tag{24}$$

The first gives the radical equation of motion,

$$\dot{P}_r = m\ddot{r} = \frac{J^2}{mr^3} \tag{25}$$

as $J = P_\theta = mr^2\dot{\theta}$, by substituting it results that

$$\dot{P}_r = \frac{m^2r^4\dot{\theta}^2}{mr^3} = mr\dot{\theta}^2 \tag{26}$$

The term $-\frac{\partial V}{\partial r}$ normally appears in this radial equation of motion and represents the force derived from a potential. In other words,

$$m\ddot{r} = mr\dot{\theta}^2 - \frac{\partial V}{\partial r} \tag{27}$$

In the present case, this term does not exist.

To complete one's knowledge of the system's development, those relations derived from the constraints, have to be resorted to.

1. $E = \frac{1}{2}m\nu^2 = const. \Rightarrow |\nu| = const$

2. $\left(\frac{d\vec{L}}{dt}\right)_{XYZ} = \left(\frac{d\vec{L}}{dt}\right)_{X'Y'Z'} + \vec{\Omega} \times \vec{L}$

where the applied external torque

$$\vec{M} = \left(\frac{d\vec{L}}{dt}\right)_{XYZ} \tag{28}$$

by hypothesis it is known that $\left(\frac{d\vec{L}}{dt}\right)_{X'Y'Z'} = \vec{0}$ and from (1) it is concluded that $\left(\frac{d\vec{L}}{dt}\right)_{X'Y'Z'} = \vec{\Omega} \times \vec{L}$
and it is obtained that $\Omega = \frac{M}{L}$.

This vector represents the rotation velocity of the frame of axes linked to the cylinder ($X'$, $Y'$, $Z'$) about the frame of inertial references axes ($X$, $Y$, $Z$).

In polar coordinates, the variable defining the rotation about the frame of axes is $\dot{\theta}$ and it is concluded that

$$\dot{\theta} = \Omega = \frac{M}{L} \tag{29}$$

substituting in the radial equation of motion, results that

$$m\ddot{r} = mr\dot{\theta}^2 = mr\Omega^2 \tag{30}$$

Moreover, $P_\theta$=const. and at the initial moment equals

$$P_\theta = mr^2\dot{\theta} = mr\nu \tag{31}$$

as $\nu$ is constant, it is concluded that $r$ is also constant. Finally it is found that

$$r = \frac{\nu}{\Omega} \qquad \text{and} \qquad m\ddot{r} = m\nu\Omega \tag{32}$$

## Further considerations

The Hamiltonian is independent of $\theta$. This is an expression of the system's rotation symmetry or, in other words, there is no preferred alignment in the plane.

The equation $\frac{\partial H}{\partial \theta} = 0$ means that the energy of the system remains invariable if it is turned to a new position without changing $r$, $P_r$ or $P_\theta$. This aspect is clearly proved.

Worthy of mention in this case is the fact that the Hamiltonian equations do not provide, in this type of problem, complete information on the radial movement. The reason for this is that the central force is a function of $\theta$ which is a coordinate that does not appear in the Hamiltonian (it can be ignored) and can only be calculated using the constraints.

This proposal is valid for any property of the particle that displays vectorial characteristics and is sensitive to an external torque of the type described, without the need to ascertain the real physical nature of that property. The proposal can be extended to equivalent problems, despite their not involving the presence of internal angular momentum as a central characteristic.

Poisson brackets. The Poisson bracket $[P_\theta, H]$ is

$$[P_\theta, H] = \left[\frac{\partial P_\theta}{\partial \theta}\frac{\partial H}{\partial P_\theta} - \frac{\partial P_\theta}{\partial P_\theta}\frac{\partial H}{\partial \theta}\right] = -\frac{\partial H}{\partial \theta}$$

as $\frac{\partial H}{\partial \theta} = 0$, it is concluded that $[P_\theta, H] = 0$.

## CHARGED PARTICLE IN A MAGNETIC FIELD

Let us apply this theory to the specific case of a charged particle with spin $\vec{s}$ and magnetic momentum $\vec{\mu}$, moving at a velocity $\vec{\nu}_0$ in a uniform magnetic field of flux density $\vec{B}$.

As it's well known, the interaction between $\vec{B}$ and $\vec{\mu}$ makes a torque to act upon the particle. This situation verifies every condition we have defined in our hypothesis, and so, the behaviour of the particle must be in agreement with the theoretical model we have developed above.

Considering the most general of the cases, let us suppose that $\vec{\mu}$ forms unknown angles $\theta$ with $\vec{B}$ and $\alpha$ with $\vec{\nu}_0$. This situation is represented in the following figure (See Fig. 2).

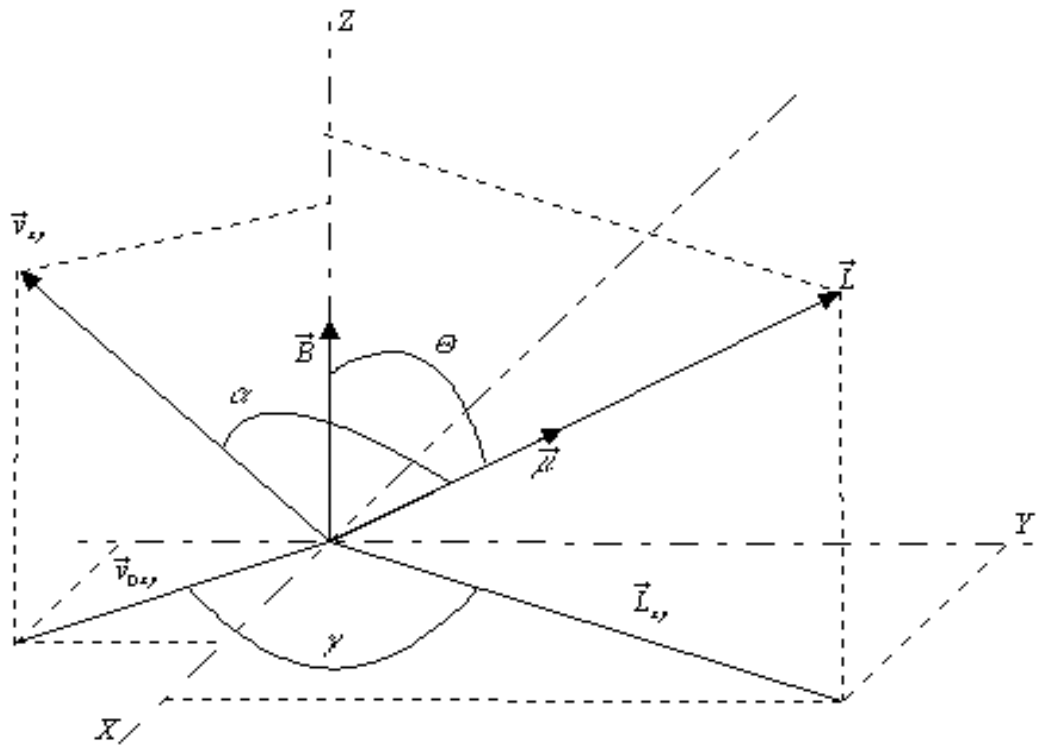

Figure 2. Charged particle in a magnetic field.

It is clear that the rotation of $\vec{\nu}_0$ causes rotation of the $\vec{\nu}_0$ component on the $XY$ plane, $\nu_{0xy}$. Furthermore, according to the theory here presented, the particle will trace a helix and the $\vec{\nu}_0$ component in the direction of the $Z$ axis, $\vec{\nu}_{0z}$, will not be affected.

In this case the angular momentum $\vec{L} = \vec{s}$.

If the particle stops inside $\vec{B}$, the rotational velocity of the angular momentum would be expressed by:

$$\Omega = \frac{|\vec{\mu} \times \vec{B}|}{|\vec{L}|} = \frac{|\vec{M}|}{|\vec{L}|} \tag{33}$$

Obviously, if the angle $\alpha$ formed by $\vec{s}$ and $\vec{\nu}_0$ remains constant, the angle $\gamma$ formed by $\vec{s}_{xy}$ and $\vec{\nu}_{0xy}$ will also remain constant. The velocity of $\vec{\nu}_{axy}$ can be calculated merely by calculating the velocity of $\vec{s}_{xy}$.

$$s_{xy} = s \sin\theta \tag{34}$$

and

$$\Omega \nu_{0xy} = \Omega s_{0xy} = \frac{|\vec{\mu}||\vec{B}|\sin\theta}{L \sin\theta} = \frac{\mu B}{L} \tag{35}$$

We can conclude from this expression that the angular velocity of the rotation of the particle depends on the magnetic momentum $\vec{\mu}$, the magnetic field $\vec{B}$ in which the particle is immersed and the spin $\vec{s}$ of the particle, and it is independent of the relative positions of $\vec{\mu}$, $\vec{B}$ and $\vec{\nu}_0$.

In the case in question, the angular velocity of the rotation of the particle in $\vec{B}$ is expressed by:

$$\vec{\Omega} = \gamma \frac{e}{2m}\vec{B} \tag{36}$$

According to the theory developed above, the particle will draw a circular trajectory. We can determine the radius of the orbit drawn by the electron under the influence of a magnetic field

$$r = \frac{\nu_0}{\Omega} = \frac{2m\nu_0}{\gamma eB} \tag{37}$$

In the specific case in which the particle is an electron, $\gamma$ has a value of 2 and the radius of the orbit is calculated from the following equation:

$$r = \frac{\nu_0}{\Omega} = \frac{m\nu_0}{eB} \tag{38}$$

From the conclusions summarized above, we state that the particle is submitted to a central force affecting the charged particle inside the magnetic field. To calculate its value, we only need to recall the formula:

$$\vec{F} = m\vec{\nu}_0 \times \vec{\Omega} \tag{39}$$

and substitute $\vec{\Omega}$ for its value previously obtained. Then we get:

$$\vec{F} = m\vec{\nu}_0 \times \gamma \frac{e}{2m}\vec{B} \tag{40}$$

Once again, if we apply this to the case in which the particle is an electron, the gyromagnetic factor is 2, and hence

$$\vec{F} = e\vec{\nu}_0 \times \vec{B} \tag{41}$$

The behaviour predicted by the theory here shown of a charged spinning particle penetrating into a magnetic field with velocity $\nu$, matches the one that can be observed in the laboratory, and the force to which it will be submitted is the well known Lorentz force.

.
.
.

.
.
.

## EXPERIMENT SUGGESTED TO TEST THIS THEORY: INTERACTION OF A HOMOGENEOUS MAGNETIC FIELD AND A PERPENDICULAR GYRATING MAGNETIC FIELD, WITH A PARTICLE WITH SPIN AND MAGNETIC MOMENT

### Formulation of the problem

The particle with spin, $\vec{S}$, moves with velocity $\vec{\nu}$. It first passes into a homogeneous magnetic field $\vec{B}'$, inducing the magnetic moment, $\vec{\mu}$, of the particle, polarized in the direction of the field. It later enters a region in which a homogeneous field, $\vec{B}$, and a gyrating field, $\vec{H}_0$, with rotation frequency, $\dot{\phi}$, are superposed. Fields $\vec{B}'$ and $\vec{B}$ are parallel and in fact could even be the same field, although their function is different in each region.

Reference frame (See Fig. 8).

$XYZ$ Fixed system in the particle but which does not rotate with this.

$\vec{B}$ is on the $Z$ axis.

$\vec{H}_0$ is on the $XY$ plane.

$X'Y'Z'$ Fixed frame in the particle which precesses with this.

$\vec{S}$ and $\vec{\mu}$ are on the $Z'$ axis.

$\vec{\nu}$ is on the $Y'$ axis.

$X_0Y_0Z_0$ Reference frame from which observations are made.

$\vec{B}$ is on the $Z_0$ axis.

$\vec{H}_0$ is on the $X_0Y_0$ plane.

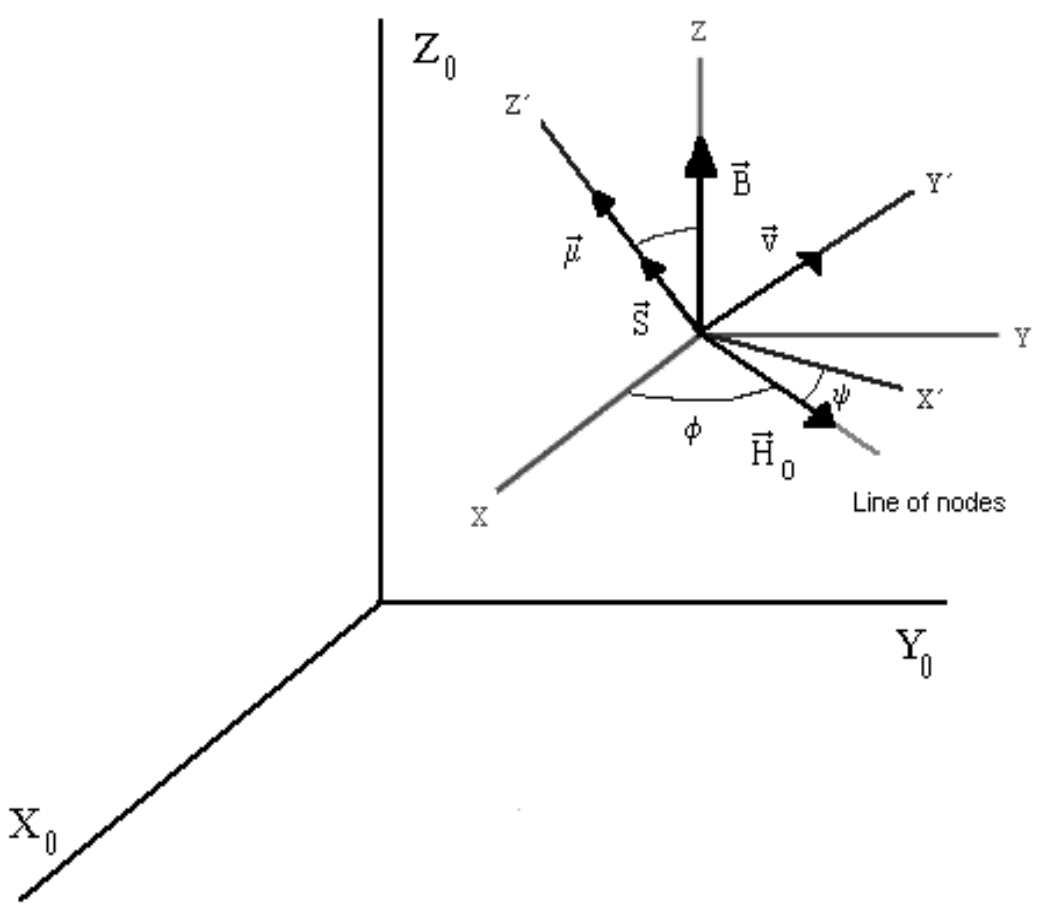

Figure 8. Reference frames.

The rotation velocity, $\vec{\Omega}$, of the frame of axes linked to the solid, $X'Y'Z'$, will be expressed within this system as:

$$\vec{\Omega} = \Omega_1\vec{i}' + \Omega_2\vec{j}' + \Omega_3\vec{k}' \tag{51}$$

This rotation velocity, $\vec{\Omega}$, can also be expressed by using Euler angles as:

$$\begin{aligned} \Omega_1 &= \dot{\phi}\sin\theta\sin\psi + \dot{\theta}\cos\psi \\ \Omega_2 &= \dot{\phi}\sin\theta\cos\psi - \dot{\theta}\sin\psi \\ \Omega_3 &= \dot{\phi}\cos\theta + \dot{\psi} \end{aligned} \tag{52}$$

Our particle can also rotate with respect to the frame of axes, $X'Y'Z'$, and the most general expression to describe this angular velocity is:

$$\vec{\omega} = \omega_{X'}\vec{i}' + \omega_{Y'}\vec{j}' + \omega_{Z'}\vec{k}' \tag{53}$$

It is proved that $\dot{\psi} = \omega_{Z'}$. The proof is easy, for if the $Z'$ axis is fixed, i.e. both $\phi$ and $\theta$ are constant, the rotations about the $Z'$ axis are rotations of the $X'$ and $Y'$ axes in the $X'Y'$ plane. Hence, the Euler angle that reports the rotation is $\psi$. It can then be concluded that

$$\dot{\psi} = \omega_{Z'} \tag{54}$$

By agreement, but without loss of generality, let us consider that the particle has an angular momentum, $\vec{S}$, that is constant in module.

## Frame of axes linked to the solid

In the definition of the frame of axes linked to the solid, we can either choose:

a) The frame of axes strictly accompanies the particle in the rotation which, in classical terms, gives its internal angular momentum.

b) The frame of axes linked to the solid is defined by a parallel axis to the internal angular momentum of the particle, and the other two are perpendicular to each other and are in a plane which is perpendicular to the angular momentum. Internal angular momentum is understood to be a quality of the particle whose mathematical characteristics coincide with those of the angular momentum, without considering the physical nature of this quality. In other words, no hypothesis is formed as to whether or not the angular momentum implies rotations of the particle under study.

For the purposes of our problem, definition (b) has been used.

However, it should be remembered that our particle can reach a rotation velocity, $\vec{\omega}$, as a result of the interactions to which it can be subjected.

## Analysis of the interactions

Given the nature of our problem, the particle penetrates a homogeneous magnetic field, $\vec{B}$, inducing the magnetic moment, $\vec{\mu}$, of the particle, polarized in the direction of the field.

Hence, when the particle penetrates the magnetic field, $\vec{B}$ (parallel to $\vec{B}'$, and can even coincide with it) and $\vec{H}_0$, the magnetic moment, $\vec{\mu}$, is only (initially) sensitive to the gyrating magnetic field, $\vec{H}_0$.

This, consequently, will be the first interaction that we analyze.

## Interaction with the magnetic field, $\vec{H}_0$

The interaction of the gyrating magnetic field, $\vec{H}_0$, with the magnetic moment, $\vec{\mu}$, does not modify the energy of the system, as both are permanently perpendicular.

The energy of the interaction is expressed

$$E_H = \vec{H}_0 \cdot \vec{\mu} = H_0 \mu \cos\frac{\pi}{2} = 0 \qquad (55)$$

The interaction between $\vec{H}_0$ and $\vec{\mu}$, is equivalent to the action of a torque, $\vec{\Gamma}_H$, on the particle, where

$$\vec{\Gamma}_H = \vec{\mu} \times \vec{H}_0 \qquad (56)$$

$\vec{\Gamma}_H$ is perpendicular to the $X'$ and $Z'$ axes, and is located on the $Y'$ axis.

By using the previously obtained Euler equations (9) and applying them to our problem:

$$\begin{aligned} \dot{\omega}_{Z'} &= \frac{M_3}{I_3} = 0 = \ddot{\psi} \\ \Omega_1 &= \frac{\Gamma_H}{S} = \frac{\mu H_0}{S} \\ \Omega_2 &= \frac{M_1}{S} = 0 \end{aligned} \qquad (57)$$

so that $\ddot{\psi} = 0$ in all cases, given that $M_3 = 0$ at all times. We can avoid $\dot{\psi}$ by making it constant in all cases, and for convenience $\psi = 0$ (See Fig. 9).

The Euler equations depending on the Euler angles (52), are simplified with the result that

$$\begin{aligned} \Omega_1 &= \dot{\theta} = \frac{\mu H_0}{S} \\ \Omega_2 &= \dot{\phi} \sin\theta = 0 \\ \Omega_3 &= \dot{\phi} \cos\theta \end{aligned} \qquad (58)$$

from which it is deduced that

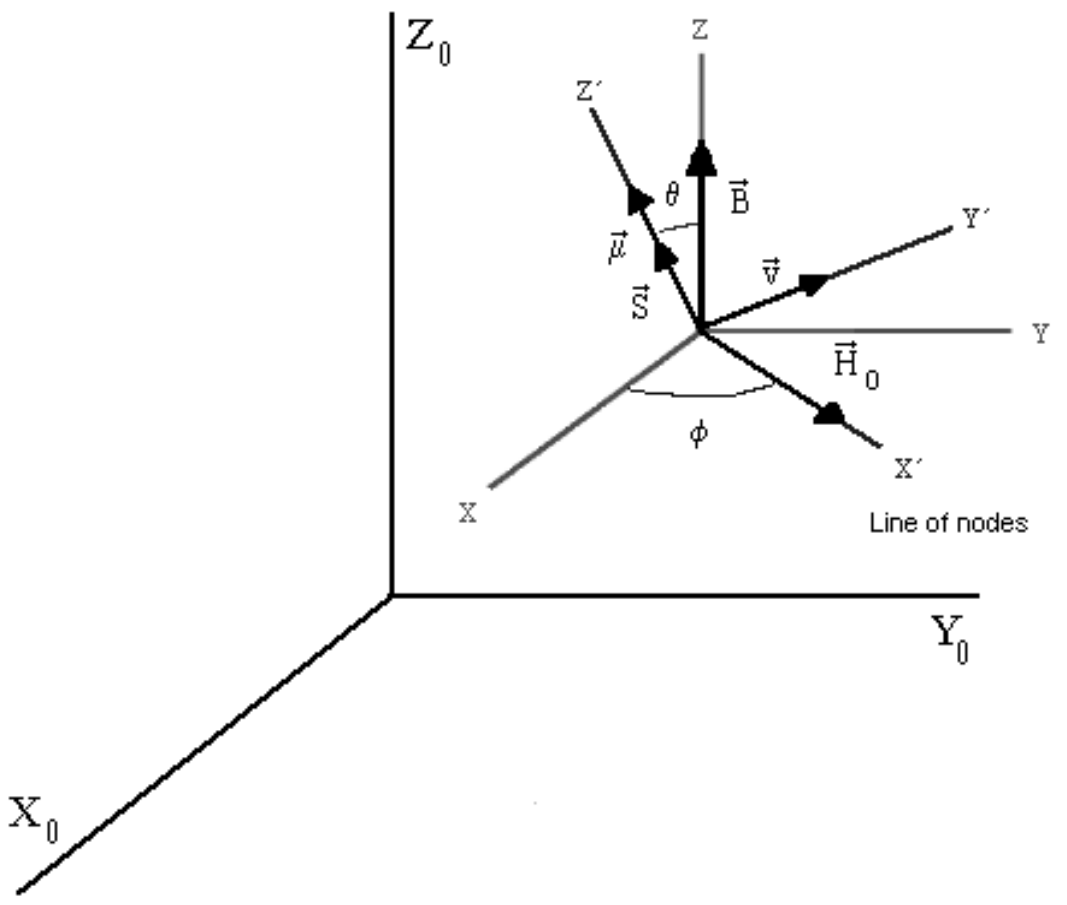

Figure 9. Interaction with the magnetic field, $\vec{H}_0$.

$$\begin{aligned} \dot{\theta} &= \frac{\mu H_0}{S} \\ \dot{\phi} &= 0 \\ \Omega_3 &= 0 \end{aligned} \qquad (59)$$

in other words, the rotation of the frame linked to the solid on $X'Y'Z'$, $\Omega_1$, is constant and equal to $\dot{\theta}$, where $\dot{\theta}$ is the rotation of the frame of axes linked to the solid with regard to $XYZ$ expressed in the Euler angles.

We have obtained the rotation velocity, $\dot{\theta}$, but we have to calculate the rotation radio $r(t)$ if we want to know the trajectory. To do this we will use the following equality between differential operators,

$$\frac{d}{dt} = \frac{d^*}{dt} + \vec{\Omega}\times \quad ,$$

where
$\frac{d}{dt}$ is the derivative in the inertial frame, in our case $X_0Y_0Z_0$,
$\frac{d^*}{dt}$ is the derivative in the non-inertial frame, in our case $X'Y'Z'$,
$\vec{\Omega}$ is the angular rotation velocity of the frame of axes linked to the solid in relation to the inertial frame of axes.

We will apply this equality to $\vec{r}$, a vector of position of one frame with respect to another, and for convenience we will express all the vectors in the $X'Y'Z'$ frame.

$$\frac{d^*\vec{r}}{dt} = \frac{d\vec{r}}{dt} - \vec{\Omega} \times \vec{r} \qquad (60)$$

If we represent the equation in components, we have

$$\begin{aligned} \dot{r}_{X'} + r_{Z'}\Omega_2 - r_{Y'}\Omega_3 &= \nu_{X\,0} \\ \dot{r}_{Y'} + r_{X'}\Omega_3 - r_{Z'}\Omega_1 &= \nu_{Y\,0} \\ \dot{r}_{Z'} + r_{Y'}\Omega_1 - r_{X'}\Omega_2 &= \nu_{Z\,0} \end{aligned} \quad (61)$$

Taking into account the initial conditions (See Fig. 10), $\Omega_2 = 0$ and $\Omega_3 = 0$.

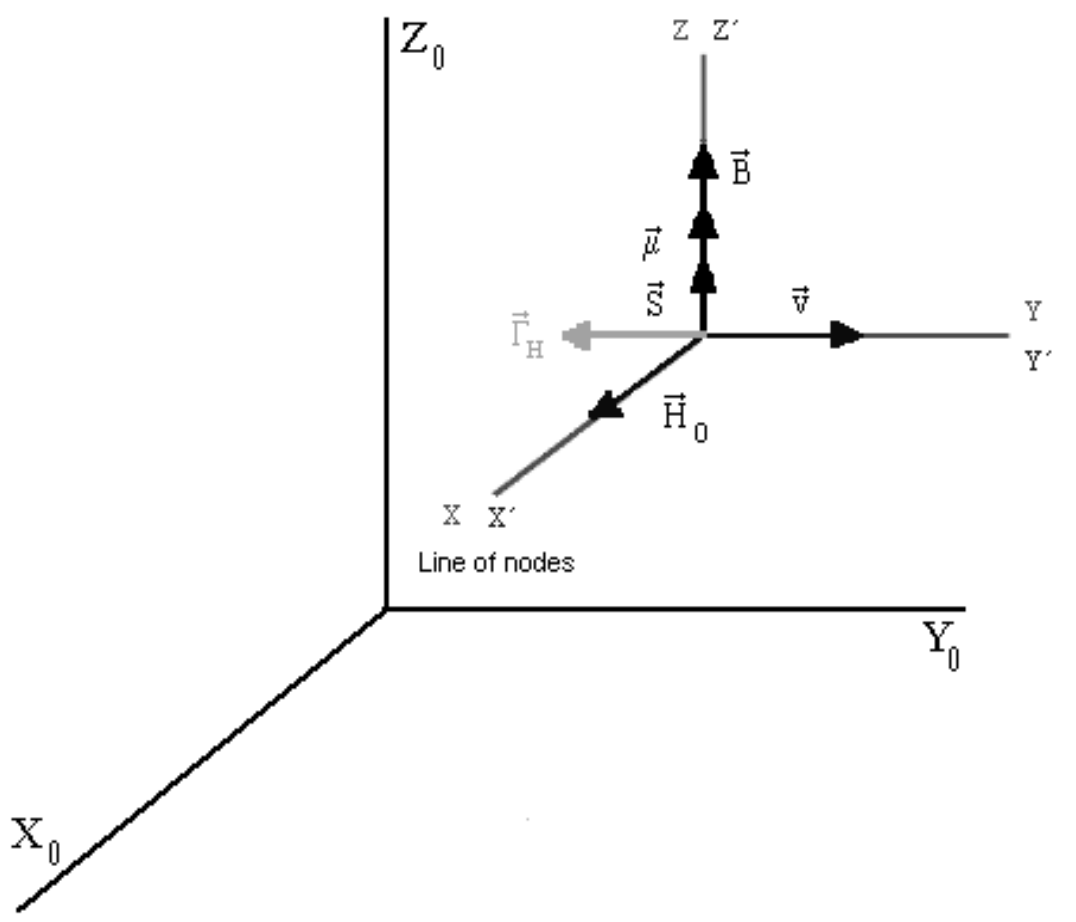

Figure 10. Initial conditions.

The interaction does not modify the energy of the system, as explained beforehand, so the module of $\vec{\nu}$ has to remain constant.

$$|\vec{\nu}| = \nu = \text{constant}.$$

and, $\nu_{X\,0} = 0$, $\nu_{Y\,0} = \nu$, $\nu_{Z\,0} = 0$.

As the energy of the system cannot vary, $\dot{r}_{X'} = \dot{r}_{Y'} = \dot{r}_{Z'} = 0$. If they differed from zero tangential accelerations would be involved and therefore a variation in the total energy of the system.

These conditions simplify the equations of our system (61) and we obtain

$$r_{Z'} = -\frac{\nu}{\Omega_1} \quad (62)$$

This is to say, the particle and the frame of axes linked to the particle describe a circular trajectory with respect to the observation frame. The movement plane is perpendicular to $\vec{H}_0$.

## Interaction with the magnetic field $\vec{B}$

At the moment the circular trajectory is commenced, $\theta$ is no longer null and the interaction with the field, $\vec{B}$, begins (See Fig. 11).

$$\begin{aligned} \vec{\Gamma} &= \vec{\mu} \times \vec{B} && (63) \\ \Omega_2 &= -\frac{\mu B \sin\theta}{S} && (64) \end{aligned}$$

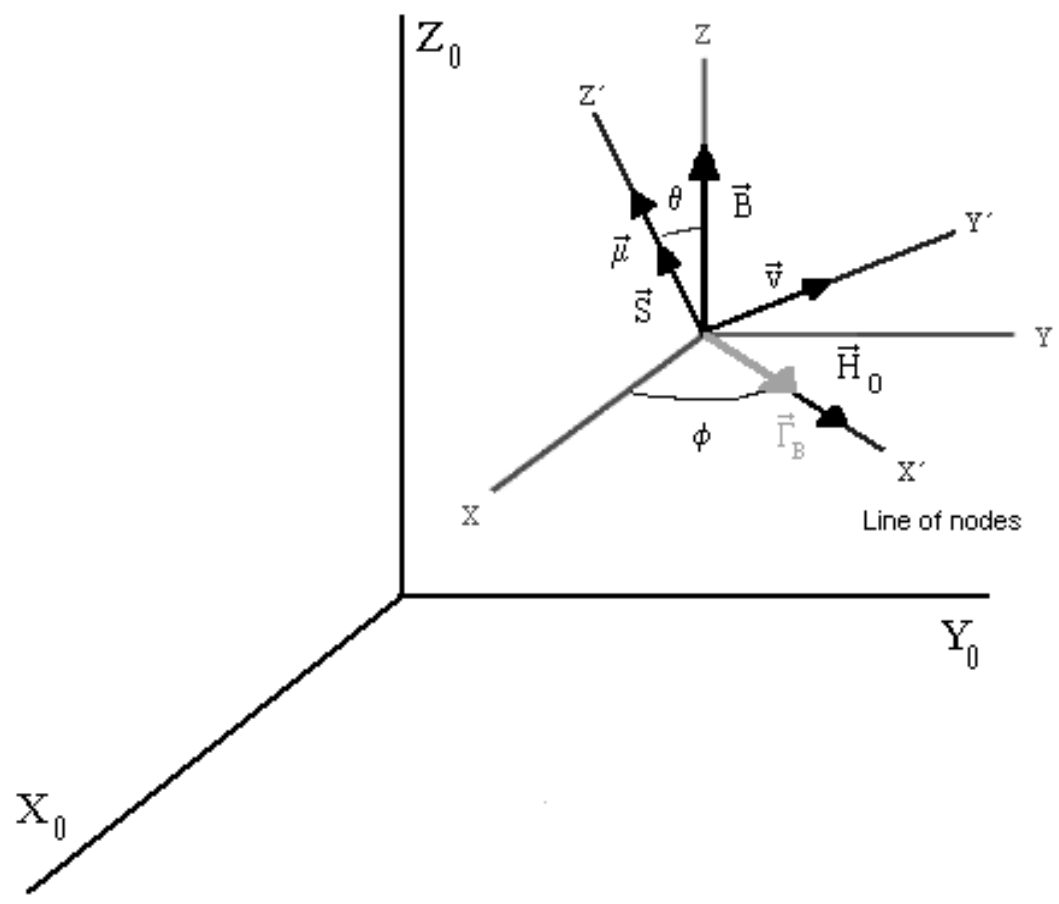

Figure 11. Interaction with the magnetic field.

$M_3$ continues to be null and, therefor, so do $\dot{\psi}$ and $\psi$, which do not vary, but $\Omega_2$ is now different from zero.

$$\begin{aligned} \Omega_1 &= \dot{\theta} \\ \Omega_2 &= \dot{\phi}\sin\theta = -\frac{\mu B \sin\theta}{S} \\ \Omega_3 &= \dot{\phi}\cos\theta \end{aligned} \quad (65)$$

We finally obtain,

$$\begin{aligned} \dot{\theta} &= \frac{\mu H_0}{S} \\ \dot{\phi} &= -\frac{\mu B}{S} \\ \Omega_3 &= \frac{\mu B \cos\theta}{S} \end{aligned} \quad (66)$$

These are the expressions of the rotation of the frame of axes linked to the solid in the base that defines the trihedral linked to the solid.

but in the base defining the trihedral of reference frame.

$$\begin{aligned} \omega_X &= \dot{\theta}\cos\phi + \dot{\psi}\sin\theta\sin\phi \\ \omega_Y &= \dot{\theta}\sin\phi - \dot{\psi}\sin\theta\cos\phi \\ \omega_Z &= \psi\cos\theta + \dot{\phi} \end{aligned} \quad (67)$$

As $\dot{\psi} = \psi = 0$, we obtain

$$\begin{aligned} \omega_X &= \dot{\theta}\cos\phi \\ \omega_Y &= \dot{\theta}\sin\phi \\ \omega_Z &= \dot{\phi} \end{aligned} \quad (68)$$

These expressions clearly reflect the behaviour of the frame of axes linked to the solid and of the solid, with respect to the reference axes.

Visualization of the movement is assisted by considering the following specific situations $\dot{\phi} = 0$

i) For $\phi = 0$

$$\begin{aligned} \omega_X &= \dot{\theta} \\ \omega_Y &= 0 \\ \omega_Z &= 0 \end{aligned}$$

We would have a rotation about the $X$ axis.

j) For $\phi = \frac{\pi}{2}$

$$\begin{aligned} \omega_X &= 0 \\ \omega_Y &= \dot{\theta} \\ \omega_Z &= 0 \end{aligned}$$

We would have a rotation about the $Y$ axis.

The rotation velocity in the $XY$ plane is

$$\begin{aligned} \vec{\omega}_{XY} &= \dot{\theta}(\cos\theta\vec{i} + \sin\theta\vec{j}) \\ |\vec{\omega}_{XY}| &= \dot{\theta} \end{aligned} \quad (69)$$

Therefore, the total movement is a circular trajectory of radius, $r_{Z'}$, perpendicular to the field, $\vec{H}_0$, which in turn spins about the field, $\vec{B}$, with velocity $\dot{\phi}$ (classically know as Larmor rotation frequency). This explains why the frequency of the oscillating field, $\vec{H}_0$, has to coincide with the Larmor frequency.

The behaviour of the particle indicates that it is subjected to two central forces

$$\vec{F}_1 = m\vec{\nu} \times \vec{\Omega}_1 = m\vec{\nu} \times \dot{\theta}\vec{i}' \quad (70)$$

$$\vec{F}_2 = m\vec{\nu} \times \dot{\phi}\vec{k} \quad (71)$$

From an energy point of view, we can confirm that the magnetic field, $\vec{H}_0$, does not modify the energy of the system, and that the magnetic field, $\vec{B}$, cyclically modifies the potencial and kinetic energy of the particle, since

$$\begin{aligned} E_B &= \vec{B}\cdot\vec{\mu}\cos\theta \\ \text{with} \qquad & \theta(t) = \frac{\mu H_0}{S}t \end{aligned} \quad (72)$$

## Additional considerations

The reader will have already noted the similarity between this behaviour and the Larmor effect.

It is know that this effect is produced by applying a magnetic field, $\vec{B}$, to a particle of charge, $q$, moving in an orbit around a fixed specific charge, $q'$. The result is a precession of the trajectory around the direction of the applied magnetic field, with a precession velocity, $\omega_L = \frac{qB}{2m}$, know as Larmor frequency, with the proviso that the cyclotronic frequency is directly obtained by this procedure.

This statement can be expressed by saying that the angular momentum vector of the particle with respect to the rotation axis, referred to as orbital angular momentum, precesses about the direction of the magnetic field, $\vec{B}$.

From our procedure, it is particularly easy to reach the same conclusions by considering that

$$\begin{aligned} \frac{d\vec{L}}{dt} &= \vec{\Gamma}_B = \vec{r}\times\vec{F} \\ \vec{L} &= \vec{r}\times m\vec{\nu} \end{aligned} \quad (73)$$

By substituting our development, $\vec{F} = m\vec{\nu}\times\vec{\Omega}$, in (73) we obtain

$$\begin{aligned} \frac{d\vec{L}}{dt} &= \vec{r}\times(m\vec{\nu}\times\vec{\Omega}) \\ &= (\vec{r}\times m\vec{\nu})\times\vec{\Omega} = \vec{L}\times\vec{\Omega} \end{aligned} \quad (74)$$

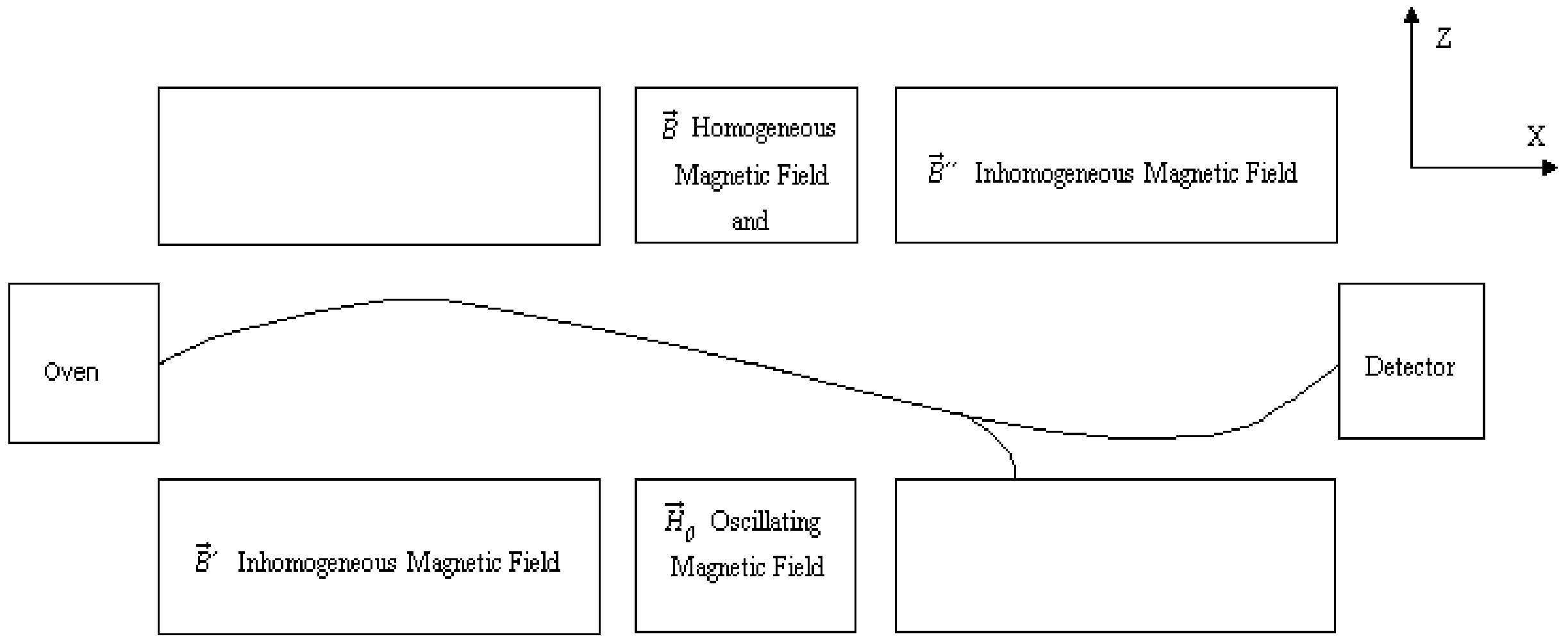

Figure 12. Rabi's experiment.

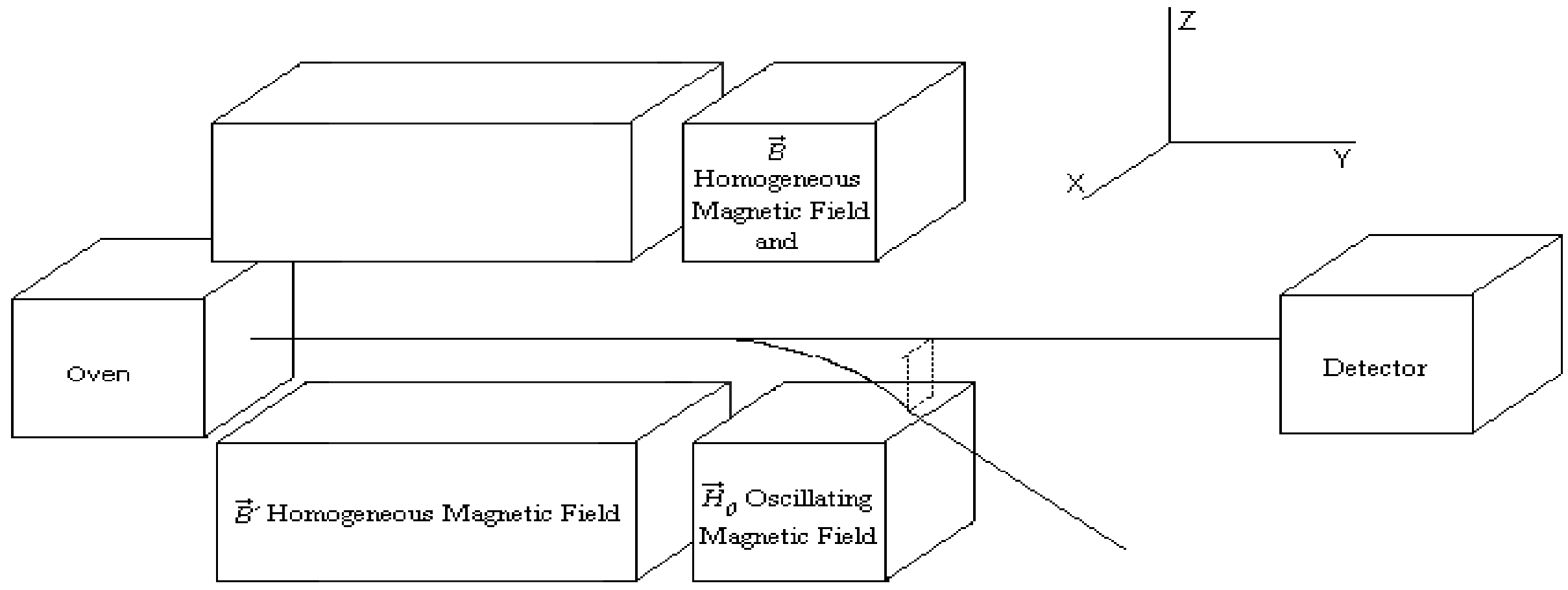

Figure 13. Experiment suggested.

(the triple vector product does not generally fulfill the associative property, but it can be demostrated in our.)

Equation (74) describes the angular momentum precession about the magnetic field with a precession rate $\vec{\Omega}$.

The Larmor effect results from subjecting the particle to the Lorentz central force, because it is within $\vec{B}$. From the following, the Lorentz force is equivalent to our force, given that

$$F = m\nu\Omega = m\nu\frac{\mu B}{S} = m\nu\frac{e}{m}B = e\nu B \qquad (75)$$

Which demostrates that both movements are equivalent.

## Experimental conditions

This experiment can be carried out by emulating the rotating magnetic field using a radio frequency magnetic field, just as Rabi did in his experiments.

Rabi's experiment (see References (2)) consists of a collimated particle beam that crosses an inhomogeneous magnetic field. It later passes through a region where a homogeneous and a radio frequency magnetic field are superposed, and finally passes through an inhomogeneous field that refocuses the beam towards the detector.

The inhomogeneous fields separate the beam into different beams according to their magnetic moment (the dependence this on the spin) as in an experiment of Stern-Gerlach. When leaving the first field, these beams later pass into the second inhomogeneous field, which refocuses the beams to the detector. By adjusting the second inhomogeneous magnetic field we will obtain refocusing conditions for a beam or group of beams.

In the central part of the experimental arrangement, the homogeneous magnetic field is superposed on the radio frequency field. In this region, the spin and magnetic moment are deflected with respect to the constant field when the oscillating field frequency approaches the Larmor precession frequency. This process is know as nuclear magnetic resonance.

After deflection, the atom is on another level which will not fulfill the refocusing condition and a decrease in the beam intensity will be observed in the detector. This procedure is used to study nuclear spin, nuclear magnetic moments and hyperfine structures. However, the behaviour of the particles in Rabi's experiment is theoretically justified in our exposition and it is therefore unnecessary to use inhomogeneous magnetic fields (as is required in Rabi's experiments).

Our experiment only requires the presence of a homogeneous magnetic field to induce the magnetic moment, $\vec{\mu}$, of the particle, polarized in the direction of the field.

In such a way that:

$1^o$ Only those particles whose Larmor frequency, $\dot{\phi}$, coincides with the radio frequency will abandon the rectilinear trajectory.

$2^o$ The trajectory described by these particles should coincide with that of the theoretical solution given in this article while they remain within the region in which the two fields coexist. According to this solution, the particle abandons the rectilinear path and the $XY$ plane.

$3^o$ When the particle abandons the region of coexisting fields, it will follow a rectilinear movement but will not reach the detector.

$4^o$ This procedure facilitates selection of the different angular moments because of their dependence on the Larmor precession.

## CONCLUSIONS

When a system with internal angular momentum, moving at constant velocity with respect to a reference frame, is subjected to an interaction of the type under study, i.e., a torque perpendicular to the internal angular momentum vector, it will begin to trace a circular path of radius

$$r = \frac{\nu}{\Omega}$$

where $\Omega = \frac{M}{L}$

Its behaviour is equivalent to that produced by subjecting the cylinder to a central force

$$\vec{F} = m\vec{\nu} \times \vec{\Omega}$$

As it can be seen in figure 14, the trajectory which particle follows is the trajectory II, while the trajectory I is the intuitive one, but as it has been shown in this paper it is not the real behaviour of the particle.

## ACKNOWLEDGMENTS

Through the development of this paper I have had the privilege of maintaining some discussions about its content with Dr. José L. Sánchez Gómez.

It will also like to thank Miguel Morales Furió for his support and help to the development of this article.

In the same way, I will like to express my gratitude to Juan Silva Trigo, Jesús Abellán, Berto González Carrera and to all the other people who has worked in the design and realization of all the experiments describe in this article, whose enumeration would be too extensive.

I would also like to express that the responsability for any affirmation here introduced lies exclusively with the author.

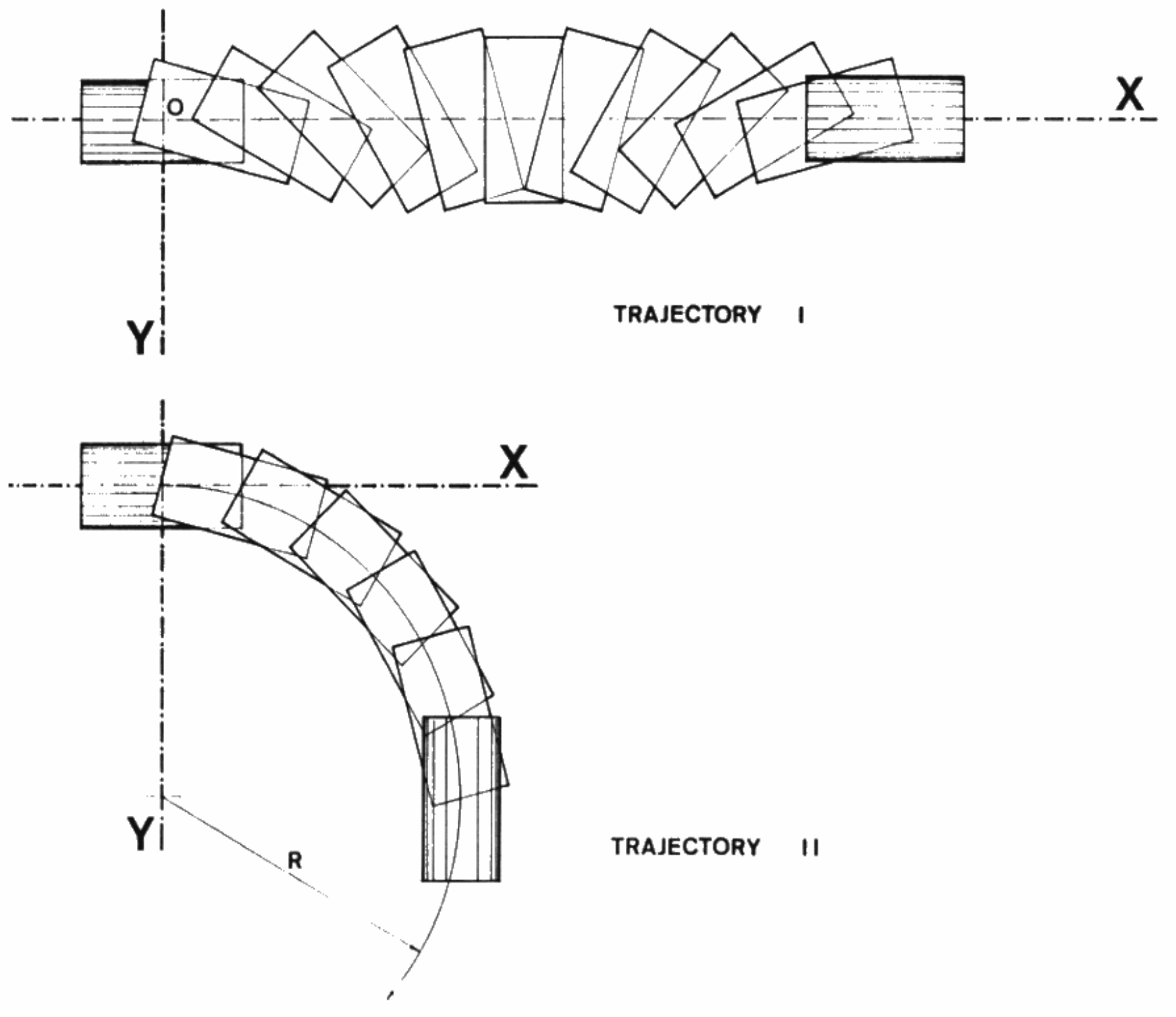

Figure 14. Classical trajectory and trajectory predicted in this article.